\documentstyle[psfig,eps]{mn}

\newif\ifAMStwofonts

\def\simgt{\hbox{\rlap{\raise 0.425ex\hbox{$>$}}\lower 0.65ex\hbox{$\sim$}}}
\def\simlt{\hbox{\rlap{\raise 0.425ex\hbox{$<$}}\lower 0.65ex\hbox{$\sim$}}}
\def\h1{$h^{-1}$}

\ifoldfss
  \ifCUPmtlplainloaded \else
    \NewTextAlphabet{textbfit} {cmbxti10} {}
    \NewTextAlphabet{textbfss} {cmssbx10} {}
    \NewMathAlphabet{mathbfit} {cmbxti10} {} 
    \NewMathAlphabet{mathbfss} {cmssbx10} {} 
  \fi
  \ifAMStwofonts
    \ifCUPmtlplainloaded \else
      \NewSymbolFont{upmath} {eurm10}
      \NewSymbolFont{AMSa} {msam10}
      \NewMathSymbol{\upi}     {0}{upmath}{19}
      \NewMathSymbol{\umu}     {0}{upmath}{16}
      \NewMathSymbol{\upartial}{0}{upmath}{40}
      \NewMathSymbol{\leqslant}{3}{AMSa}{36}
      \NewMathSymbol{\geqslant}{3}{AMSa}{3E}

    \fi
  \fi
\fi 

\ifnfssone
  \newmathalphabet{\mathit}
  \addtoversion{normal}{\mathit}{cmr}{m}{it}
  \addtoversion{bold}{\mathit}{cmr}{bx}{it}
  \newmathalphabet{\mathbfit} 
  \addtoversion{normal}{\mathbfit}{cmr}{bx}{it}
  \addtoversion{bold}{\mathbfit}{cmr}{bx}{it}
  \newmathalphabet{\mathbfss} 
  \addtoversion{normal}{\mathbfss}{cmss}{bx}{n}
  \addtoversion{bold}{\mathbfss}{cmss}{bx}{n}
  \ifAMStwofonts
    \ifCUPmtlplainloaded \else
      \UseAMStwoboldmath
      \makeatletter
      \new@mathgroup\upmath@group
      \define@mathgroup\mv@normal\upmath@group{eur}{m}{n}
      \define@mathgroup\mv@bold\upmath@group{eur}{b}{n}
      \edef\UPM{\hexnumber\upmath@group}
      \new@mathgroup\amsa@group
      \define@mathgroup\mv@normal\amsa@group{msa}{m}{n}
      \define@mathgroup\mv@bold\amsa@group{msa}{m}{n}
      \edef\AMSa{\hexnumber\amsa@group}
      \makeatother
      \mathchardef\upi="0\UPM19
      \mathchardef\umu="0\UPM16
      \mathchardef\upartial="0\UPM40
      \mathchardef\leqslant="3\AMSa36
      \mathchardef\geqslant="3\AMSa3E
    \fi
  \fi
\fi 
\ifnfsstwo
  \DeclareMathAlphabet{\mathbfit}{OT1}{cmr}{bx}{it}
  \SetMathAlphabet\mathbfit{bold}{OT1}{cmr}{bx}{it}
  \DeclareMathAlphabet{\mathbfss}{OT1}{cmss}{bx}{n}
  \SetMathAlphabet\mathbfss{bold}{OT1}{cmss}{bx}{n}
  \ifAMStwofonts
    \ifCUPmtlplainloaded \else
      \DeclareSymbolFont{UPM}{U}{eur}{m}{n}
      \SetSymbolFont{UPM}{bold}{U}{eur}{b}{n}
      \DeclareSymbolFont{AMSa}{U}{msa}{m}{n}
      \DeclareMathSymbol{\upi}{0}{UPM}{"19}
      \DeclareMathSymbol{\umu}{0}{UPM}{"16}
      \DeclareMathSymbol{\upartial}{0}{UPM}{"40}
      \DeclareMathSymbol{\leqslant}{3}{AMSa}{"36}
      \DeclareMathSymbol{\geqslant}{3}{AMSa}{"3E}
    \fi
  \fi
\fi 

\ifCUPmtlplainloaded \else
  \ifAMStwofonts \else 
    \def\upi{\pi}
    \def\umu{\mu}
    \def\upartial{\partial}
  \fi
\fi

\title[Galactic Contribution in Red Quasars]{Host Galaxy Contribution to the
Colours of `Red' Quasars}
\author[F. J. Masci et al.]
{F. J. Masci,$^1$\thanks{Email: {\bf \tt fmasci@physics.unimelb.edu.au}}
R. L. Webster$^1$\thanks{Email: {\bf \tt rwebster@physics.unimelb.edu.au}}
and P. J. Francis$^2$\thanks{Email: {\bf \tt pfrancis@mso.anu.edu.au}}
\\$^1$School of Physics, University of Melbourne,
\\~Parkville, Victoria 3052, Australia
\\$^2$Department of Physics and Theoretical Physics, Faculty of Science,
\\~Australian National University, Canberra, ACT 0200, Australia}
\date{Zeroth Draft} 
\pubyear{1998}

\begin{document}

\maketitle

\newcommand{\fmmm}[1]{\mbox{$#1$}}
\newcommand{\scnd}{\mbox{\fmmm{''}\hskip-0.3em .}}
\newcommand{\scnp}{\mbox{\fmmm{''}}}

\begin{abstract}
We describe an algorithm that measures self-consistently 
the relative galaxy contribution
in a sample of radio-quasars
from their optical spectra alone. 
This is based on a spectral fitting method 
which uses the size of the characteristic 4000\AA~ feature of elliptical
galaxy SEDs. 
We apply this method to the Parkes Half-Jansky Flat Spectrum
sample of Drinkwater et al. (1997) to determine whether emission from
the host galaxy can significantly contribute to the very red
optical--to--near-infrared colours observed.
We find that at around $2\sigma$ confidence, most of the reddening in
unresolved (mostly quasar-like) sources 
is unlikely to be due to contamination by a red stellar component. 
\end{abstract}


\section{Introduction}

The recent discovery of large 
numbers of radio-selected quasars with 
very red optical--to--near-infrared colours 
suggests that existing quasar surveys may be severely incomplete. 
Based on a high identification rate in the optical and near-infrared,
Webster et al. (1995) found a broad range of colours with
$2<B-K<10$ for flat-spectrum radio quasars in a subsample of the
Parkes 2.7GHz survey (Drinkwater et al. 1997; hereafter `Parkes quasars'). 
For comparison, quasars selected by standard optical techniques
show a small scatter around
$B-K\sim2.5$.
Two theories have been proposed to explain the large scatter: 
first, Webster et al. (1995) interpreted this in terms of 
extinction by line-of-sight dust (see also Masci 1997), 
and second, Serjeant \& Rawlings (1996)
suggested that this was due to intrinsically red 
optical/infrared synchrotron radiation associated with the radio emission.  
This paper explores a third possibility: that the red colours
are due to contamination by starlight from the host galaxies. 

In a recent near-infrared imaging study of a sample of radio-quasars selected from
the low-frequency (408MHz) catalog, Benn et al. (1998) found that sources
with red $B-K$ colours to $z\sim2$ could be explained by an excess of 
host galaxy light in $K$. Most of the images appeared non-stellar (or resolved) 
suggesting that indeed starlight was responsible for 
the redness in $B-K$ colour.
It is important to note however that all these sources were associated with
extended radio galaxies of which a majority are of the steep-spectrum type, a
common feature of low-frequency selected samples. A large fraction
are also often associated with luminous giant ellipticals. 
In view of the simple orientation-based 
unified model for radio-loud AGN, it is possible that
galaxy light is more easily detected in these sources due to anisotropic 
obscuration and/or unbeamed emission of the central AGN. 
The radio-quasars explored in this paper are all of the flat-spectrum type
selected at moderately high frequency. Thus, they are not expected to exhibit 
similar properties in the near-infrared. 

The $B-K$ colours of
normal radio 
galaxies are known
to be quite red, exhibiting a similar dispersion to those observed in
the Parkes sample.
These sources are often associated with giant ellipticals and
their colours 
appear to be uniformly distributed over the range
$3\simlt B-K\simlt 7$ for redshifts $z\simlt2$~\cite{McCarthy1993}.
It is possible that the host galaxies of Parkes quasars
also exhibit similar properties.
To determine the importance of
host galaxy light in Parkes 
quasars,
we need to quantify its contribution
throughout the optical to near-IR wavelength region.
 
Determining the host galaxy properties of radio-quasars and BL-Lac-type 
sources are crucial
for studies of AGN evolution and testing unified schemes.
A significant number of BL-Lacs and other core-dominated radio loud AGN 
are surrounded by nebulosities whose optical spectra 
are very similar to those of giant ellipticals
\cite{MFH1978,Ulrich1988,Taylor1996}. 
A majority of these
host galaxies
have been detected in sources at relatively low redshifts,
$z\simlt0.1$, where the galaxy flux is easily detected. 
Since most quasars are at high redshifts, host galaxy detections 
have been
difficult due to contamination by their strong nuclear emission.
Recent high resolution imaging using HST
has revealed that a majority of high redshift quasars 
also reside in giant ellipticals 
\cite{Hutchings1995,Bahcall1995}.

The optical spectral energy distributions (SEDs) 
of ellipticals are all similar in shape.
At $\lambda\simgt3000$\AA, they rise steeply up to $\sim4000$\AA$\,$
where a sharp step-like cutoff is observed. This cutoff is often
refered to as the ``4000 Angstrom break'' and is 
caused by the dearth of hot and young, usually O and B-type stars
with time, whose spectra mostly dominate at wavelengths from
$\approx 4000$\AA~ into the UV.
For $\lambda\simlt4000$\AA, 
the existence of a multiplet of heavy metal absorption lines arising from
stellar atmospheres, 
results in a steepening of the break. 
The presence of this feature thus provides a signature
for determining whether 
an underlying elliptical host is contributing to the total light.
 
In this paper, we describe a new unbiased
method that uses the size of this 
characteristic break in the optical spectra of Parkes quasars 
to quantify the host galaxy contribution.
We use this method to investigate whether the host galaxies of Parkes
quasars can significantly
contribute to the optical and near-IR continua and hence 
cause the spread in $B-K$ colours observed.
 
This paper is organised as follows:
In section~\ref{outgc}, we outline our method used to quantify the host galaxy
contribution. A discussion of the spectral data 
and assumed input parameters
is given in section~\ref{data}.
Results are presented in section~\ref{regcp}.
Implications for AGN unified schemes are explored in section~\ref{tuf} and
results discussed in section~\ref{diseven}.
All results are summarised in section~\ref{concsix}.

\section{Outline of Method}
\label{outgc}
 
The galaxy contribution can be estimated using an
algorithm that determines the relative strength of the 
characteristic 4000\AA$\,$ break in the optical spectra of Parkes sources
in an unbiased way.
No other feature in a generic quasar spectrum will mimic this feature.
In this section, we describe this algorithm.
 
\subsection{Assumptions}
 
There are two important assumptions that will be required by our
algorithm. 
First, we assume that the spectral shape defining an underlying quasar
optical continuum is `smooth' and contains no breaks. 
This is justified by our current knowledge of quasar optical spectra.
Our choice for this shape is somewhat arbitrary, and will
be further discussed in section~\ref{method}.  

\begin{figure}
\vspace{-1in}
\plotonesmall{1.1}{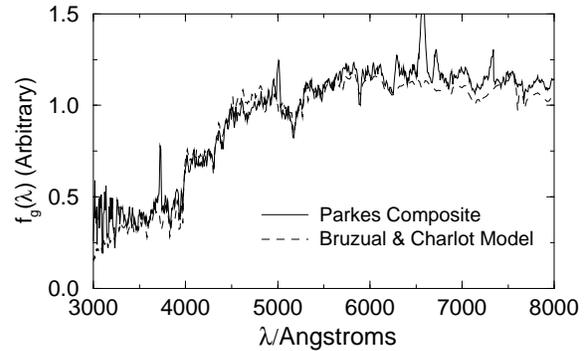}
\vspace{-1.6in}
\caption[]{Composite optical spectrum of 18 Parkes sources which appear
spatially extended in $B_{J}$ (solid curve), and a model spectrum for an `old'
elliptical SED from Bruzual \& Charlot (1993) (dashed curve).}
\label{galspec}
\end{figure}

Second, we need to assume
the input galaxy spectrum which defines the shape of the 4000\AA$\,$
break to be used in our analysis.
We have formed a composite optical spectrum of 18 Parkes sources all of which
appear `spatially extended' in $B_{J}$.
This composite is shown as the solid curve in Fig.~\ref{galspec} and
clearly shows the characteristic 4000\AA$\,$ break.
Superimposed (dashed) is an elliptical galaxy SED predicted from the
stellar population synthesis models of 
Bruzual \& Charlot~\shortcite{Bruzual1993}.
This model is for a galaxy of age 8Gyr, and 
assumes a 1Gyr burst described by a Salpeter initial
mass function with no star formation
thereafter. 
In fact, any `old aged' model with long completed star formation
will in general be similar. 
Models assuming `long lived' constant star formation rates
lead to considerably bluer optical-UV continua, characteristic 
of those observed in 
spirals.
The composite and model in Fig.~\ref{galspec}
are in excellent agreement, except for the
emission lines arising from ISM gas in the composite.
In our analysis, we shall assume that this model represents the 
underlying elliptical SED in each of our Parkes sources. 
 
\begin{figure}
\vspace{-1in}
\hspace{-0.3in}
\plotonesmall{1.1}{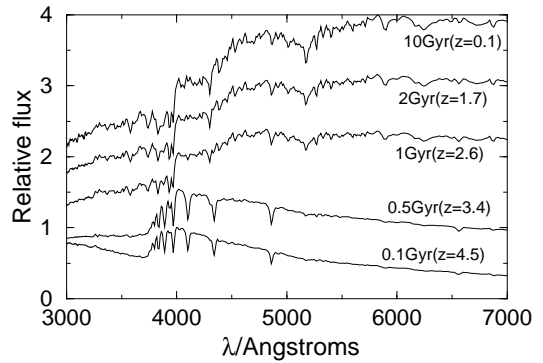}
\vspace{-1.5in}
\caption[]
{Spectral evolution of the 1Gyr burst model of Bruzual \& Charlot (1993).
The age in Gyr and the approximate corresponding
redshift for a formation epoch $z_{f}=5$ with
($q_{0},h_{50}$)=(0.5,1) is shown beside each spectrum.}
\label{specevol}
\end{figure}

For sources with redshifts $z\simgt1$, the $4000$\AA$\,$ break is redshifted 
out of the wavelength range available in the spectra of Parkes quasars 
(see section~\ref{apal}) and thus, cannot be used in our algorithm.
We are therefore restricted to $z<1$.
For simplicity, we assume that the general shape of our model 
elliptical SED is independent of redshift to $z\sim1$.
The colours of radio galaxies observed to $z\sim3$ are consistent
with formation redshifts $4\simlt z_{f}\simlt 20$ 
\cite{Spinrad1987,Dunlop1989}.
These formation redshifts are also suggested by models of
galaxy formation in the CDM scenario~\cite{White1991}.
The spectral synthesis models  
predict that the general form of an elliptical SED shown in 
Fig.~\ref{galspec} can be immediately established following an
almost instantaneous ($\sim1$Gyr) burst of star formation. 
Adopting the 1Gyr burst model of Bruzual \& Charlot~\shortcite{Bruzual1993} 
(updated
1995 models), we show in Fig.~\ref{specevol} the spectral evolution as a
function of age since the initial starburst.
Also shown beside each age is the approximate
redshift for a formation epoch of $z_{f}=5$.
Initially, the 4000\AA$\,$ break region is diluted by starburst activity.
After a few gigayears,
the stellar population evolves passively, maintaining 
an almost uniform SED shape (see Bruzual \& Charlot 1993). 
For formation epochs $z\simgt5$, 
it is apparent that the generic elliptical SED shape of Fig.~\ref{galspec} 
is easily established by $z\sim1$. 

\subsection{The Algorithm}
\label{method}
 
In general, the total flux at a given wavelength observed in a Parkes source,
$f_{T}(\lambda)$, can be modelled as the sum of light contributed
by the central quasar or AGN, $f_{q}(\lambda)$, and any underlying host
galaxy $f_{g}(\lambda)$ (eg. Fig.~\ref{galspec}). 
We write the relationship 
between these quantities as:
\begin{equation}
f_{q}(\lambda)\,=\,f_{T}(\lambda) - c\,f_{g}(\lambda), 
\label{fq}
\end{equation} 
where $c$ is a scaling factor giving an arbitrary measure of the
amount of galaxy light we wish to determine for a particular source. 
 
Given $f_{T}(\lambda)$ for a particular source and an
arbitrary galaxy spectrum $f_{g}(\lambda)$ (Fig.~\ref{galspec}), our aim is to 
determine the value of $c$ such that $f_{q}(\lambda)$ looks 
something like a quasar spectrum. 
From our knowledge of quasar optical spectra, 
an obvious choice is to require that $f_{q}(\lambda)$ be ``smooth'' and
contain no breaks. 
Thus, the basis of this algorithm involves subtracting 
an arbitrary amount
of galaxy flux, $c\,f_{g}(\lambda)$ from $f_{T}(\lambda)$ such that  
the resulting spectrum $f_{q}(\lambda)$ appears smooth (see below).
When this is achieved, the fraction of total light at a given wavelength
contributed by the host galaxy can be estimated by normalising:
\begin{equation}
F_{gal}(\lambda)\,=\,\frac{c\,f_{g}(\lambda)}{f_{T}(\lambda)}. 
\label{galfrac}
\end{equation}

In order to implement the above algorithm, we need to define
an acceptable form for the shape of the quasar spectrum $f_{q}(\lambda)$. 
Our only requirement is that this spectrum be smooth and hence our 
choice is somewhat arbitrary.
We choose $f_{q}(\lambda)$ to be a power-law, parameterised by:
\begin{equation}
f_{q}(\lambda)\,=\,f_{q}(\lambda_{max})\left(\frac{\lambda}
{\lambda_{max}}\right)^{\alpha}, 
\label{pl2}
\end{equation} 
where the slope $\alpha$ is determined between two fixed wavelengths,
$\lambda_{min}$ and $\lambda_{max}$ (see below):
\begin{equation}
\alpha\,=\,\frac{\ln\left[f_{q}(\lambda_{min})/f_{q}(\lambda_{max})\right]}
{\ln\left[\lambda_{min}/\lambda_{max}\right]}.
\label{alpha2}
\end{equation} 
Using Eqn.~\ref{fq}, the fluxes defined in Eqn.~\ref{alpha2} can be written: 
\begin{equation}
f_{q}(\lambda_{min})\,=\,f_{T}(\lambda_{min}) - c\,f_{g}(\lambda_{min})
\label{fqminmax}
\end{equation}
\[
f_{q}(\lambda_{max})\,=\,f_{T}(\lambda_{max}) - c\,f_{g}(\lambda_{max}).
\]
For a discussion on how the wavelengths 
$\lambda_{min}$ and $\lambda_{max}$ are chosen 
and the fluxes in Eqn.~\ref{fqminmax} measured, see section~\ref{apal}. 
The validity of our assumption of a single power-law for $f_{q}(\lambda)$ 
is discussed in section~\ref{diseven}. 

\begin{figure}
\vspace{0.1in}
\plotonesmall{1}{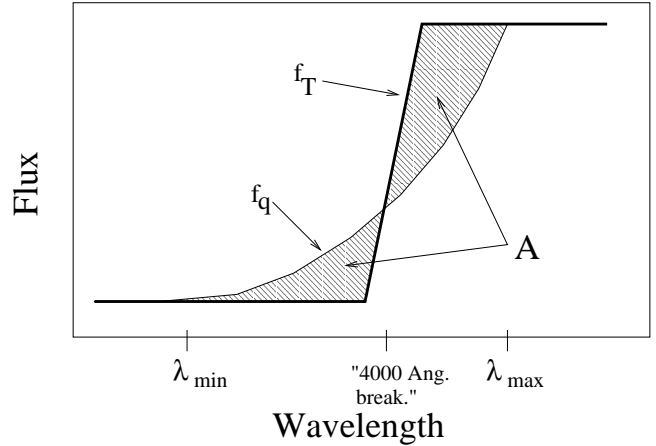}
\caption[]{Definition of our ``figure of merit'' $A$.
For a particular source spectrum $f_{T}(\lambda)$ (whose 4000\AA$\,$ break
region is shown exagerated), we subtract an amount of galaxy
flux, $f_{g}(\lambda)$ (Fig.~\ref{galspec}) until $A$ is a
minimum and a
smooth power-law $f_{q}(\lambda)$ results.
}
\label{schematic}
\end{figure}

To apply this algorithm in a self consistent way to each of our
optical spectra, we need to define a figure of merit indicating 
the point at which the maximum amount of galaxy spectrum has been 
subtracted and a ``smooth'' quasar spectrum (ie. a power-law) is
achieved.  
Let us first consider the rest frame optical spectrum of a source,
$f_{T}(\lambda)$,  
suspected of containing a 4000\AA$\,$ break. This is illustrated in  
Fig.~\ref{schematic}.
Furthermore, let us consider a ``smooth'' power-law (ie. 
the underlying quasar spectrum $f_{q}(\lambda)$)
between two wavelengths
$\lambda_{min}$ and $\lambda_{max}$ on either side of the 
4000\AA$\,$ break region as shown. 
We define our figure of merit as representing the
area $A$ of the shaded region in Fig.~\ref{schematic}.
As certain amounts of galaxy spectrum,
$c\,f_{g}(\lambda)$ (where $f_{g}(\lambda)$ is given by Fig.~\ref{galspec}) 
are gradually 
subtracted from $f_{T}(\lambda)$, $A$ will decrease
and becomes a minimum when the break disappears.
We can thus determine the value of $c$ when this occurs, allowing us
to estimate the fractional galaxy contribution from Eqn.~\ref{galfrac}.
 
For a given amount of
subtracted galaxy flux, $c\,f_{g}(\lambda)$,
we can write $A$ in terms of $c$ from Eqn.~\ref{fq} as follows:
\begin{equation}
A\,=\,\int^{\lambda_{max}}_{\lambda_{min}}
{\left| f_{q}(\lambda) - f_{T}(\lambda) + c\,f_{g}(\lambda)\right|}
\,\,d\lambda. 
\label{Aone}
\end{equation}
With $f_{q}(\lambda)$ defined by Eqns.~\ref{pl2},~\ref{alpha2} and 
~\ref{fqminmax}, $A$ can be written:
\begin{eqnarray}
\nonumber
A\,& = & \,\int^{\lambda_{max}}_{\lambda_{min}}
\put(0,20){\line(0,-1){35}}\left[f_{T}(\lambda_{max})\,-\,c\,f_{g}(\lambda_{max}
)\right]
\left(\frac{\lambda}{\lambda_{max}}\right)^{\alpha (c,f_{T},f_{g})}\\
& &{\rm\hspace{0.5in}}\,-\,f_{T}(\lambda)\,
+\, c\,f_{g}(\lambda){\rm\hspace{0.05in}}\put(0,20){\line(0,-1){35}}\,\,
d\lambda,
\label{Atwo}
\end{eqnarray}
where $\alpha(c,f_{T},f_{g})$ is defined by Eqns.~\ref{alpha2} and 
~\ref{fqminmax}. 
Thus, one would only have to minimise $A$ with respect to $c$ in order to
determine the maximal galactic contribution. 
We introduce however,  
an additional factor in Eqn.~\ref{Atwo} whose purpose 
will be to make best use of the 
available data 
and optimise our algorithm.

\subsubsection{Data Optimisation}
 
Due to systematic effects, each Parkes optical spectrum suffers from a
considerable amount of noise at the range of observed wavelengths.
Although the noise may have some wavelength dependence, we assume it is
constant. 
Given this assumption, we have used an optimal method that gives
more weight to those wavelengths in a particular spectrum where the residual 
$\left| f_{q}(\lambda)-f_{T}(\lambda)\right|$ (see Fig.~\ref{schematic})
is likely to be a maximum.
In other words, the 
galaxy subtraction process will depend most sensitively on the
spectral shape around the 4000\AA$\,$ break region.
Our assumption that the noise is wavelength independent will greatly 
simplify our optimisation method,
since otherwise, 
we would have to simultaneously optimise
those observed spectral regions with
the highest signal-to-noise.

The basis of this method simply involves convolving the integrand in 
Eqn.~\ref{Atwo} with a weighting function $G(\lambda)$ that gives
more weight to regions on either side of the $4000$\AA$\,$ break 
within $\lambda_{min}<\lambda<\lambda_{max}$ (see Fig.~\ref{schematic}).
We choose to define $G(\lambda)$ purely from the galaxy spectrum
$f_{g}(\lambda)$ (dashed curve in Fig.~\ref{galspec}).
This is defined as the difference
(or residual) between a power-law and the galaxy spectrum
within the range
$\lambda_{min}<\lambda<\lambda_{max}$, containing the $4000$\AA$\,$ break:
\begin{equation}
G(\lambda)\,=\,\left|f_{gPL}(\lambda) - f_{g}(\lambda)\right|, 
\label{G}
\end{equation} 
where 
$$
f_{gPL}(\lambda)\,=\,f_{g}(\lambda_{max})\left(
\frac{\lambda}{\lambda_{max}}\right)^{\alpha_{g}}
$$
and
$$
\alpha_{g}\,=\,\frac{\ln\left[f_{g}(\lambda_{min})/f_{g}(\lambda_{max})\right]}
{\ln\left[\lambda_{min}/\lambda_{max}\right]}.
$$
The function $G(\lambda)$ will peak
at wavelengths on either side of the
$4000$\AA$\,$ break. Thus, by convolving $G(\lambda)$ with Eqn.~\ref{Atwo}, 
relatively more weight will be given to spectral data at these 
wavelengths, where our algorithm mostly sensitively depends.

\begin{figure*}
\vspace{-1.5in}
\plotonesmall{1.3}{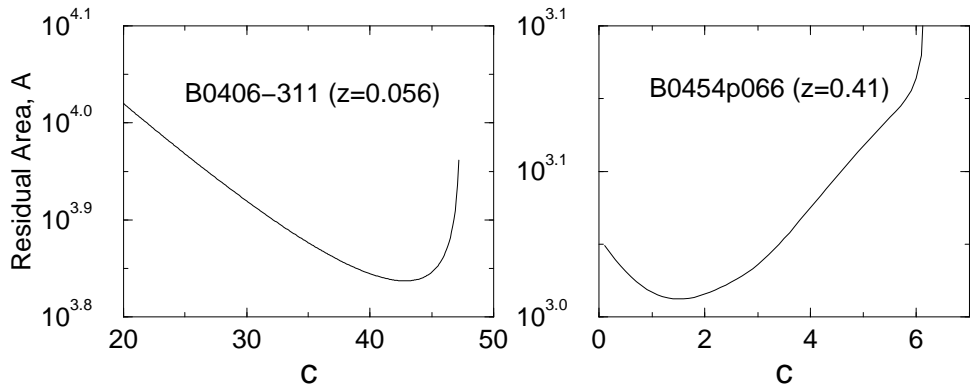}
\vspace{-1.9in}
\caption[]{
Examples of our figure of merit function $A$ (defined by
Eqn.~\ref{Athree})
as a
function of the parameter $c$. Source spectra are shown in Fig.~\ref{fits1}.}
\label{mineg}
\end{figure*}

Thus, the function to minimise can now be written:
\begin{eqnarray}
\nonumber
A\,& = & \,\int^{\lambda_{max}}_{\lambda_{min}}
\put(0,20){\line(0,-1){35}}\left[f_{T}(\lambda_{max})\,-\,c\,f_{g}(\lambda_{max})\right]
\left(\frac{\lambda}{\lambda_{max}}\right)^{\alpha (c,f_{T},f_{g})}\\ 
& &{\rm\hspace{0.5in}}\,-\,f_{T}(\lambda)\, 
+\, c\,f_{g}(\lambda){\rm\hspace{0.05in}}\put(0,20){\line(0,-1){35}}\,\,
\,G(\lambda)\,\,d\lambda,
\label{Athree}
\end{eqnarray}
where $G(\lambda)$ is defined by Eqn.~\ref{G}.
Our aim is to determine the
value of $c$ for a particular source spectrum $f_{T}(\lambda)$ 
such that $A$ is a minimum.                          
In Fig.~\ref{mineg}, we show examples of the figure of merit function $A$
(Eqn.~\ref{Athree}) as a function of the parameter $c$ for two sources
whose spectra are shown in Fig.~\ref{fits1}.
If the weighting factor $G(\lambda)$ in 
Eqn.~\ref{Athree} is neglected, we find that the values of $c$ at which 
$A$ is a minimum decrease by up to $5\%$ in most sources
where the break feature appears relatively weak or absent.
In sources with strong 4000\AA$\,$ breaks, there is no 
significant difference.
Given the value of $c$ that minimises $A$, the fractional 
galaxy contribution at some wavelength can now be computed using 
Eqn.~\ref{galfrac}.

\subsection{Error Determination}
\label{errdes}
 
In spectra where the galaxy contribution is relatively 
weak, an upper limit on its contribution at some appropriate
level of significance
would be required.
In order to do so, both random and systematic errors need to be investigated.
The magnitude of these two types of errors are estimated and compared in 
section~\ref{sys.vs.ran}. 
Here, we briefly outline the method in their determination.
 
\subsubsection{Random (Statistical) Errors}
 
The random error at some confidence level 
in the galaxy contribution is estimated by computing the value of $c$
corresponding to the statistical error in $A$ (Eqn.~\ref{Athree}).
Since Eqn.~\ref{Athree} is actually a discrete sum over wavelength bins
$\lambda_{i}$, from $\lambda_{min}$ to $\lambda_{max}$,
the statistical error is determined by adding the error for each
individual bin in quadrature, so that
\begin{equation}
\sigma(A)\,=\,\sqrt{\sum_{\lambda_{i}=\lambda_{min}}^{\lambda_{max}}
\sigma^{2}_{i}(I)}, 
\label{rand}
\end{equation}
where $\sigma_{i}(I)$ is the error in the integrand $I$ of 
Eqn.~\ref{Athree} for bin $\lambda_{i}$. 
$\sigma_{i}(I)$ will depend only on uncertainties in the measured source fluxes
$f_{T}(\lambda_{max})$ and $f_{T}(\lambda_{min})$ 
in Eqn.~\ref{Athree}.
As shall be discussed in section~\ref{apal}, these fluxes are
estimated by calculating the median continuum flux in
wavelength bins centered on $\lambda_{max}$ and $\lambda_{min}$. 
We estimate the corresponding uncertainties by computing the
rms deviation from the mean spectral flux in these wavelength bins. 

\subsubsection{Systematic Errors}
 
Our definition of the underlying quasar continuum $f_{q}(\lambda)$,
in each source only requires that it be smooth and contain no breaks.
This implies that the shape of $f_{q}(\lambda)$ is somewhat arbitrary
and thus it is possible that our quantitative measures of the galaxy 
contribution may strongly depend on its assumption in our algorithm.
The uncertainty introduced by this possible 
systematic effect will be investigated. 
 
All our calculations assume that $f_{q}(\lambda)$
is a power-law (ie. Eqn.~\ref{pl2}), since the continua of optical quasar
spectra are well represented by $f_{\nu}\propto\nu^{-\alpha}$ where
$\alpha\sim0.2-0.3$~\cite{Francis1996}.
To investigate the effects of assuming a 
different form for $f_{q}(\lambda)$ however,
we also apply our algorithm by assuming for simplicity that
$f_{q}(\lambda)$ is a straight line joining $\lambda_{min}$ and
$\lambda_{max}$ in Fig.~\ref{schematic}.
This is parameterised as follows:
\begin{equation}
f_{q}(\lambda)_{line}=\frac{\left[f_{q}(\lambda_{max}) - 
f_{q}(\lambda_{min})\right]}{\lambda_{max} - \lambda_{min}}
(\lambda - \lambda_{max})\,+\,f_{q}(\lambda_{max}), 
\label{sl}
\end{equation}
where $f_{q}(\lambda_{min})$ and $f_{q}(\lambda_{max})$ 
are defined by Eqn.~\ref{fqminmax}.
A comparison in the fractional galaxy contributions
resulting from our use of a power-law (Eqn.~\ref{pl2}) 
and a straight line (Eqn.~\ref{sl})
for $f_{q}(\lambda)$ in our algorithm, will enable us to estimate
the magnitude of this systematic effect. 
Results are presented in 
section~\ref{sys.vs.ran}.
Other possible sources of systematic error are also discussed in this
section.

\subsection{Summary}
 
To summarise, we have presented in this section a method to determine
the relative galaxy contribution in each Parkes source in a robust 
way. 
Our algorithm requires the following two input assumptions:
First, an elliptical optical SED, $f_{g}(\lambda)$, defining 
the shape of the characteristic 4000\AA$\,$ break. This we assume is a model
from Bruzual \& Charlot (1993) (see Fig.~\ref{galspec}). Second,
we require a spectral shape 
defining the underlying quasar spectrum $f_{q}(\lambda)$.
Unless otherwise specified, all our calculations shall assume a power-law
for $f_{q}(\lambda)$.
 
With the only requirement that $f_{q}(\lambda)$ be smooth and contain
no breaks, the ``suspected'' 4000\AA$\,$ break in each source spectrum
$f_{T}(\lambda)$, is subtracted until the residual between
$f_{q}(\lambda)$ and galaxy subtracted source spectrum is a minimum.
The galaxy contribution is estimated from the amount of galaxy, 
$f_{g}(\lambda)$ subtracted.
To apply this procedure in a self-consistent manner,
we have defined a figure of merit given by Eqn.~\ref{Athree}.
This is minimised with respect to the parameter $c$ from which the
fractional galaxy contribution can be easily computed using Eqn.~\ref{galfrac}.  
\section{Data and Input Parameters}
\label{data}
 
Out of the 323 sources in the Drinkwater et al.~\shortcite{Drinkwater1997} 
sample,
we have optical spectra for 194 or 60\% of the sample.
Some of these are from the compilation of 
Wilkes et al.~\shortcite{Wilkes1983}
and others are from recent observations on the AAT 
and ANU 2.3m (see Drinkwater et al. 1997).
For these latter observations (65 sources), 
the spectra cover the observed wavelength range:
$3200{\rm \AA}\simlt\lambda\simlt10000{\rm \AA}$ 
with a resolution of $\sim5$\AA$\,$ 
in the blue and $\sim20$\AA$\,$ in the red ($\simgt5200$\AA). 
Other spectra have 
typically a mean resolution
$\sim10$\AA$\,$ and cover the range: 
$3200{\rm \AA}\simlt\lambda\simlt8000{\rm \AA}$.

\subsection{Applying the Algorithm}
\label{apal}
 
Before applying our algorithm, each Parkes optical spectrum is redshifted
to its rest frame.
We then define the fixed rest frame wavelengths $\lambda_{min}$ and
$\lambda_{max}$ within which the suspected 4000\AA$\,$ break and our
figure of merit (Eqn.~\ref{Athree}) is defined (see Fig.~\ref{schematic}).
This wavelength region is chosen using the following criteria:
first, the highest redshift sources
will have the rest wavelength $\lambda_{max}$
redshifted out of the observational wavelength range of the spectra.
These sources will not be able to be used in our algorithm.
We therefore need to choose $\lambda_{max}$
such that the number of sources in which our algorithm
can be applied is not significantly reduced.
Second, we need a wavelength range
$\lambda_{min}<\lambda<\lambda_{max}$
that makes `optimal' use of 
the shape of the 4000\AA$\,$ break
region defining our figure of merit $A$ 
(the residual area in Fig.~\ref{schematic}).
In other words, we need to ensure that this region
is unambiguously defined and clearly represented in each source spectrum. 
As a compromise, we assume 
$\lambda_{min}=3500$\AA$\,$ and $\lambda_{max}=5080$\AA$\,$ in every source. 

Relative measures of the fluxes: $f_{T}(\lambda_{min,max})$ and
$f_{g}(\lambda_{min,max})$ in the source and galaxy rest frame spectra
respectively (see Eqn.~\ref{Athree}), are determined as follows.
We first define 
wavelength bins of width $\sim200$\AA$\,$ and
$\sim400$\AA$\,$ centered on $\lambda_{min}$
and $\lambda_{max}$ respectively, and then calculate the median
{\it continuum} flux in each bin.
The wavelength regions defining these bins however may
contain absorption
and emission lines. 
Such lines
are likely to bias
our estimates of the continuum level in these regions.
From the bin widths defined above, the rest wavelength regions
of interest 
are:  
$3400{\rm\AA}<\lambda<3600$\AA$\,$ and $4880{\rm\AA}<\lambda<5280$\AA$\,$. 
From the available source spectra, we find that 
no lines are likely to 
contaminate the short wavelength bin. 
For the long wavelength bin however, we find that the emission line
doublet [OIII]$\lambda\lambda$4959,5007 and a weak absorption 
feature at $\sim5170$\AA$\,$ (possibly from MgI) are strong contaminants.
To avoid significant contamination, our algorithm excludes regions of width 
10\AA$\,$ centered on these lines. 
 
Given the definitions above, the rest wavelength range required by our
algorithm will 
be $3400{\rm \AA}\simlt\lambda_{rest}\simlt5280{\rm \AA}$.
With a maximum observed wavelength of
$\lambda_{max}({\rm obs})\simeq10000$\AA$\,$
in $\sim30\%$ of the available spectra, we find that only sources
with redshifts $z\simlt0.9$ can be used
in our algorithm.
In total, we have about 53 spectra in which our algorithm can be applied.

\section{Results}
\label{regcp}
 
\subsection{Spectral Fits}
 
With the input parameters from the previous section, 
Eqn.~\ref{Athree} is minimised numerically
with respect to the parameter $c$ for each spectrum.
Having found the value $c_{min}$ that minimises Eqn.~\ref{Athree},
we can reconstruct the initial source spectrum $f_{T}(\lambda)$
around the 4000\AA~ break region within 
$3500{\rm\AA}(\lambda_{min})<\lambda<5080{\rm\AA}(\lambda_{max})$.
This is done by superimposing the maximum amount of galaxy spectrum
$c_{min}f_{g}(\lambda)$ generated by the algorithm, and a power-law
representing the underlying ``smooth'' quasar spectrum $f_{q}(\lambda)$. 
From Eqns.~\ref{fq} and~\ref{pl2},
these reconstructed model spectra can be represented:
\begin{eqnarray}
\nonumber
f_{T}(\lambda)_{model}\,& = &\,c_{min}f_{g}(\lambda)\,+\,f_{q}(\lambda)\\
\nonumber
&\equiv & c_{min}f_{g}(\lambda)\,+\,\left[f_{T}(\lambda_{max})-c_{min}\,f_{g}(\lambda_{max})\right]\\
{\rm\hspace{20mm}}& &\times\left(\frac{\lambda}{\lambda_{max}}\right)
^{\alpha (c_{min},f_{T},f_{g})}.
\label{fTmodel}
\end{eqnarray}
 
\begin{figure*}
\centerline{
\psfig{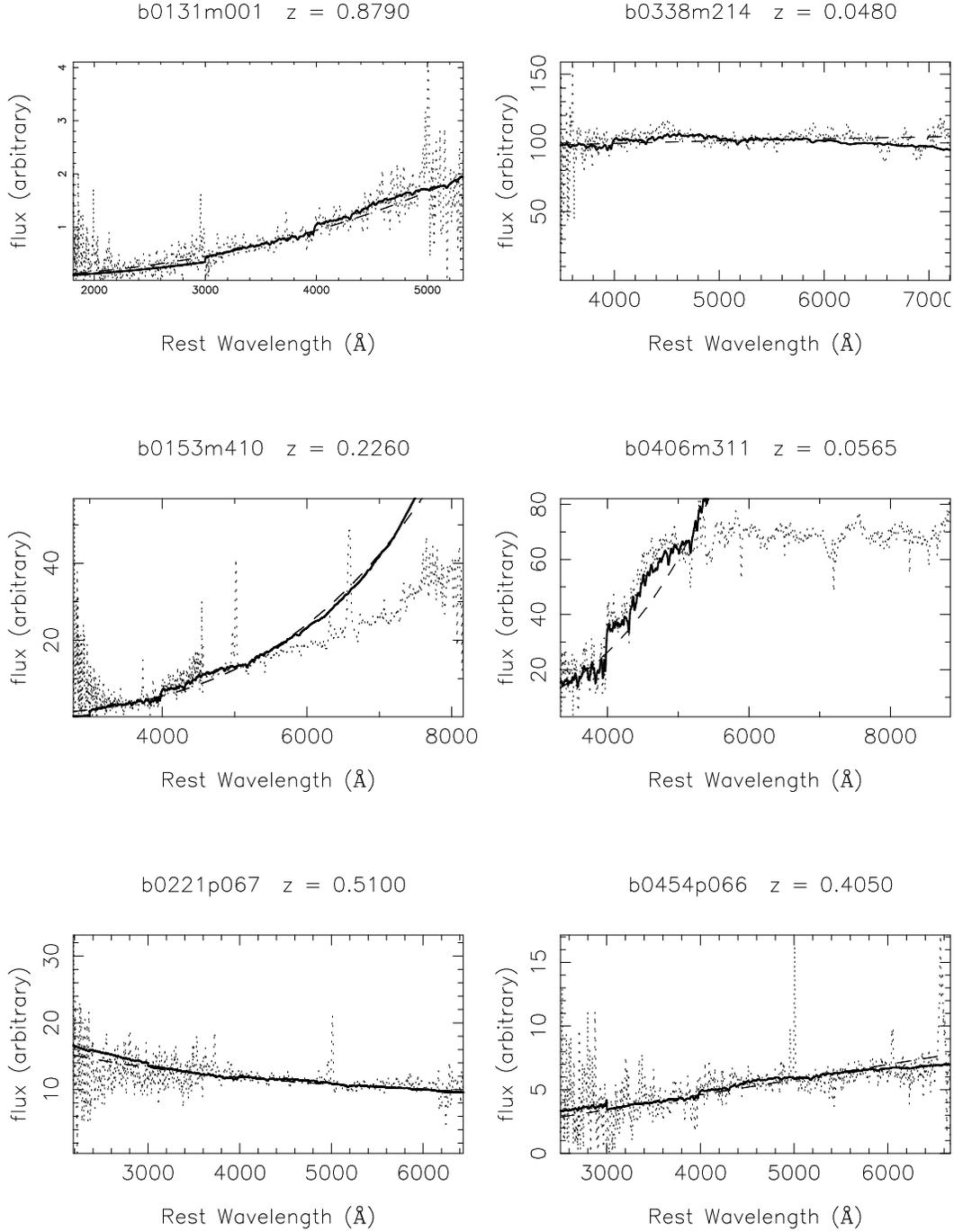}
}
\caption[]{Observed rest frame spectra (dotted) for a number of
sources in the Parkes sample. Also shown are
reconstructed model spectra (solid) as defined by Eqn.~\ref{fTmodel} and
power-law fits representing the underlying quasar continuum (dashed).
Since this quasar spectrum is purely defined within the range
$3500{\rm\AA}(\lambda_{min})<\lambda<5080{\rm\AA}(\lambda_{max})$,
the model spectra are only effective in reproducing those regions observed
within this wavelength range.}
\label{fits1}
\end{figure*}

A visual comparison between these model and observed spectra in the range
$\lambda_{min}<\lambda<\lambda_{max}$ will allow us to investigate the
accuracy of our algorithm in reproducing the observed spectra
around the 4000\AA$\,$ break region.
Reconstructed spectra $f_{T}(\lambda)_{model}$ are compared with the observed 
spectra, $f_{T}(\lambda)$, for a number of sources in Fig.~\ref{fits1}.
On each spectrum, we also show our power-law fit $f_{q}(\lambda)$.
Since the power-law spectral index for 
$f_{q}(\lambda)$ is defined 
purely within $\lambda_{min}<\lambda<\lambda_{max}$ 
(as required by our algorithm; see Eqns~\ref{pl2} and~\ref{alpha2}), 
a model spectrum will attempt 
to accurately reproduce that observed region which falls 
in this wavelength range only.  
As expected, observed spectra showing strong 4000\AA$\,$ breaks with SEDs
similar to that given in Fig.~\ref{galspec} are reproduced very accurately
about the break region.

In a majority of observed spectra where no significantly strong
breaks are discernable to the human eye however, our algorithm
nevertheless attempts to fit for a break.
Unbiased estimates of the galaxy contribution using the
relative sizes of these breaks, however weak, are presented in
section~\ref{galcont}.
 
\subsection{Systematic versus Random Errors}
\label{sys.vs.ran}

\begin{figure}
\vspace{-0.7in}
\plotonesmall{1.1}{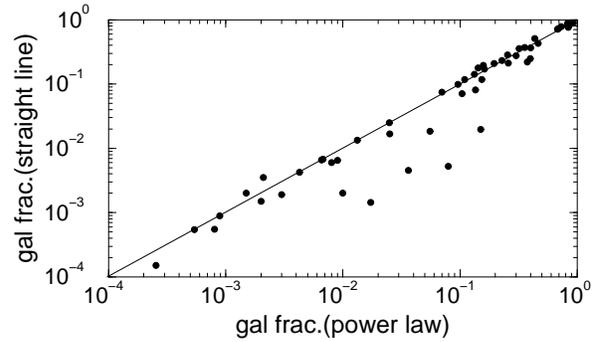}
\vspace{-2.2in}
\caption[]{
Fractional galaxy contribution at 5000\AA$\,$ (rest frame)
assuming a straight line
for $f_{q}(\lambda)$ (vertical axis) and a power-law (horizontal axis).
The diagonal line is the line of equality. (see section~\ref{sys.vs.ran}).}
\label{sys}
\end{figure}

As discussed in section~\ref{errdes}, 
a possible source of systematic uncertainty
is in our assumption of the shape of the underlying smooth quasar
continuum $f_{q}(\lambda)$.
To explore this, we compare the relative galaxy contribution obtained 
by assuming first, a power-law (PL) (Eqn.~\ref{pl2}) for $f_{q}(\lambda)$
and second, a straight line (L) (Eqn.~\ref{sl}).
Using our algorithm and these two
definitions for $f_{q}(\lambda)$, we have 
computed the fractional galaxy contribution
at 5000\AA$\,$ (rest frame). Results are shown in Fig.~\ref{sys}.
 
At first glance, estimates for the galaxy fraction using the power-law
and straight line for $f_{q}(\lambda)$ agree very well.
There is relatively little scatter about the diagonal line defining
the equality ${\rm frac}_{L}={\rm frac}_{PL}$, except for
a distinct population with ${\rm frac}_{L}<{\rm frac}_{PL}$.
No distinguishing feature in the optical spectra of this latter
class is immediately apparent.
It is likely that 
a power-law (rather than a straight line) 
within $\lambda_{min}<\lambda<\lambda_{max}$ for these sources
provides a better representation of our figure of merit $A$ 
in Fig.~\ref{schematic}. 
We quantify the systematic error from the RMS scatter in the
difference:
$\delta={\rm frac}_{L}-{\rm frac}_{PL}$, which we denote by $\sigma(\delta)$.
For this systematic effect, we therefore estimate a 
$1\sigma$ uncertainty in the galaxy fraction 
at 5000\AA$\,$ of at most 
$\sigma_{frac.}\simeq0.02$.

Another possible source of systematic error is
in our selection of the wavelengths $\lambda_{max}$ and $\lambda_{min}$,
within which the suspected 4000\AA$\,$ break and our figure of merit
(Eqn.~\ref{Athree}) is defined
(see Fig.~\ref{schematic}).
As discussed in section~\ref{apal}, the values 
$\lambda_{min}=3500$\AA$\,$ and $\lambda_{max}=5080$\AA$\,$ 
were chosen as a compromise between: 
first, to maximise the number
of sources in which the
4000\AA$\,$ break remains within the
observed wavelength after redshifting, and second, to 
make optimal use of the break region. 
What are the effects on the galaxy fraction if 
a different wavelength range were assumed?
 
To explore this, we choose to vary
$\lambda_{max}$ alone. 
Due to the
relatively small wavelength range at
$\lambda<4000$\AA$\,$ available in the galaxy spectrum (Fig.~\ref{galspec}), 
we are not as flexible in
varying $\lambda_{min}$.
We therefore keep $\lambda_{min}$ fixed at 3500\AA.
Assuming the same bin widths (200\AA$\,$ and 400\AA$\,$) 
centered on $\lambda_{min}$ and $\lambda_{max}$,
and the PL definition for $f_{q}(\lambda)$,
we have computed
galaxy fractions at 5000\AA$\,$ with 
$\lambda_{max}=5500$\AA$\,$, and 
$\lambda_{max}=6500$\AA$\,$. 
Combined with our estimates using
$\lambda_{max}=5080$\AA$\,$, we find that changing $\lambda_{max}$ makes 
neglible difference in the galaxy fraction. The fractions differ 
by no more than 1\%.

\begin{figure}
\vspace{-1.3in}
\plotonesmall{1.1}{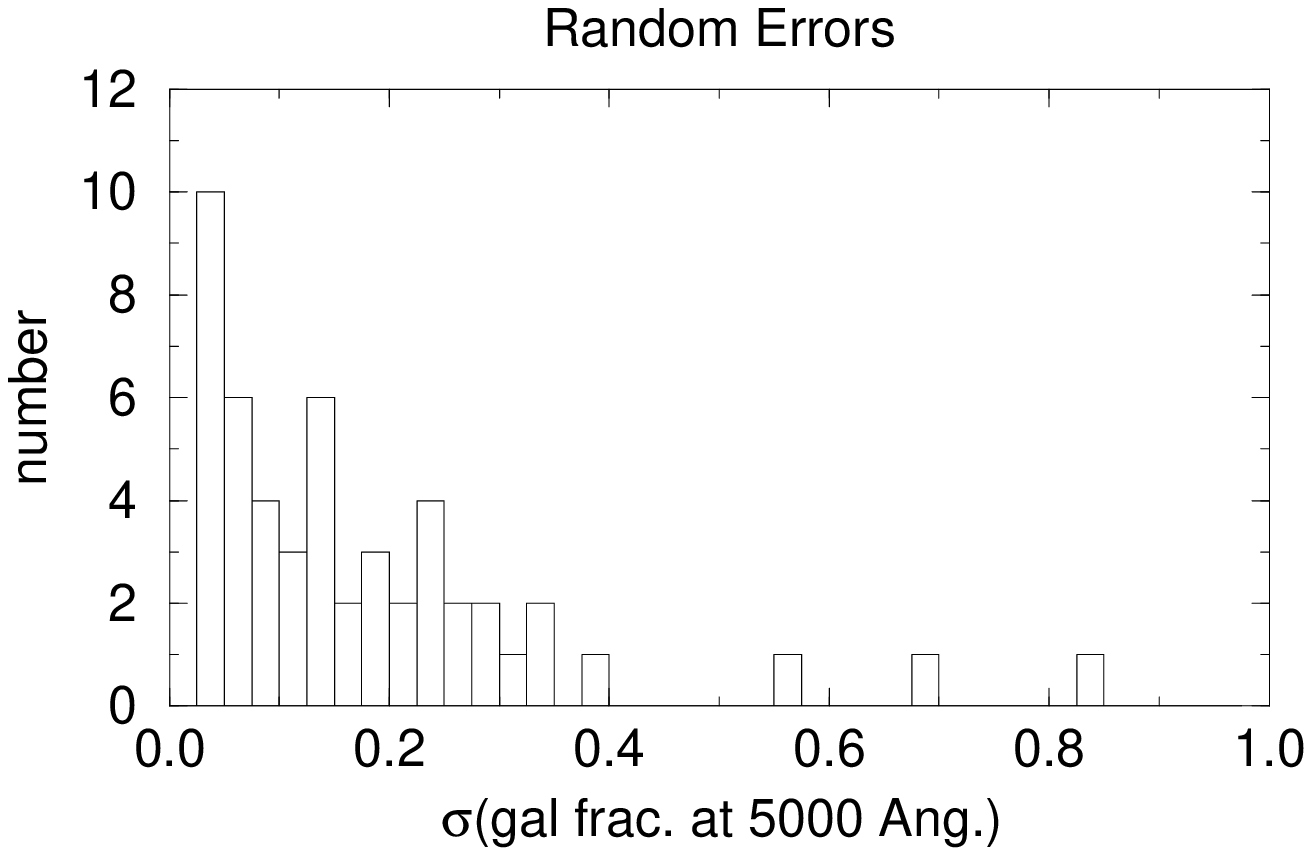}
\vspace{-1.7in}
\caption[]
{Distribution of $1\sigma$ random errors in the
fractional galaxy contribution at 5000\AA~(rest frame)
for Parkes sources.
(see section~\ref{sys.vs.ran}).}
\label{randfig}
\end{figure}

We now compare these systematic uncertainties with estimates of
the random errors.
Fig.~\ref{randfig} shows 
the distribution of random errors in the fractional galactic contribution
at 5000\AA$\,$ as determined from our algorithm (see section~\ref{errdes}).
The range in random errors is significantly broad, with a majority
of values exceeding our maximum $1\sigma$ systematic 
uncertainty of 0.02 deduced from Fig.~\ref{sys}. 
In the remaining sections, we therefore quote all 
uncertainties in the relative galaxy contribution
as purely statistical, based
on random errors alone.
 
\subsection{Galactic Contribution to the Observed Optical-Near IR
Continuum.}
\label{galcont}
 
Using our algorithm, we have computed the fractional galaxy
contribution in the individual bandpasses $B_{J}$ 
$(\lambda\simeq4400{\rm \AA})$ and $K$ $(\lambda\simeq2.2\mu{\rm m})$,
expected in an observer's frame for each source. 
We estimate these using Eqn.~\ref{galfrac} where fluxes are
approximated by computing the median spectral flux at the central wavelength
of each bandpass. 
Since our observed spectra do not extend to near-IR wavelengths,
we estimate the observed spectral flux in $K$ using our observed
$B_{J}-K$ colours and extrapolate from 
the spectral flux corresponding to $B_{J}$. 
Unfortunately, not all of the 53 sources used in
our algorithm have a measured $K$ magnitude and hence $B_{J}-K$ colour.
Of these sources, exactly 34 have known $K$ magnitudes.

\begin{figure}
\vspace{-0.2in}
\plotonesmall{1.1}{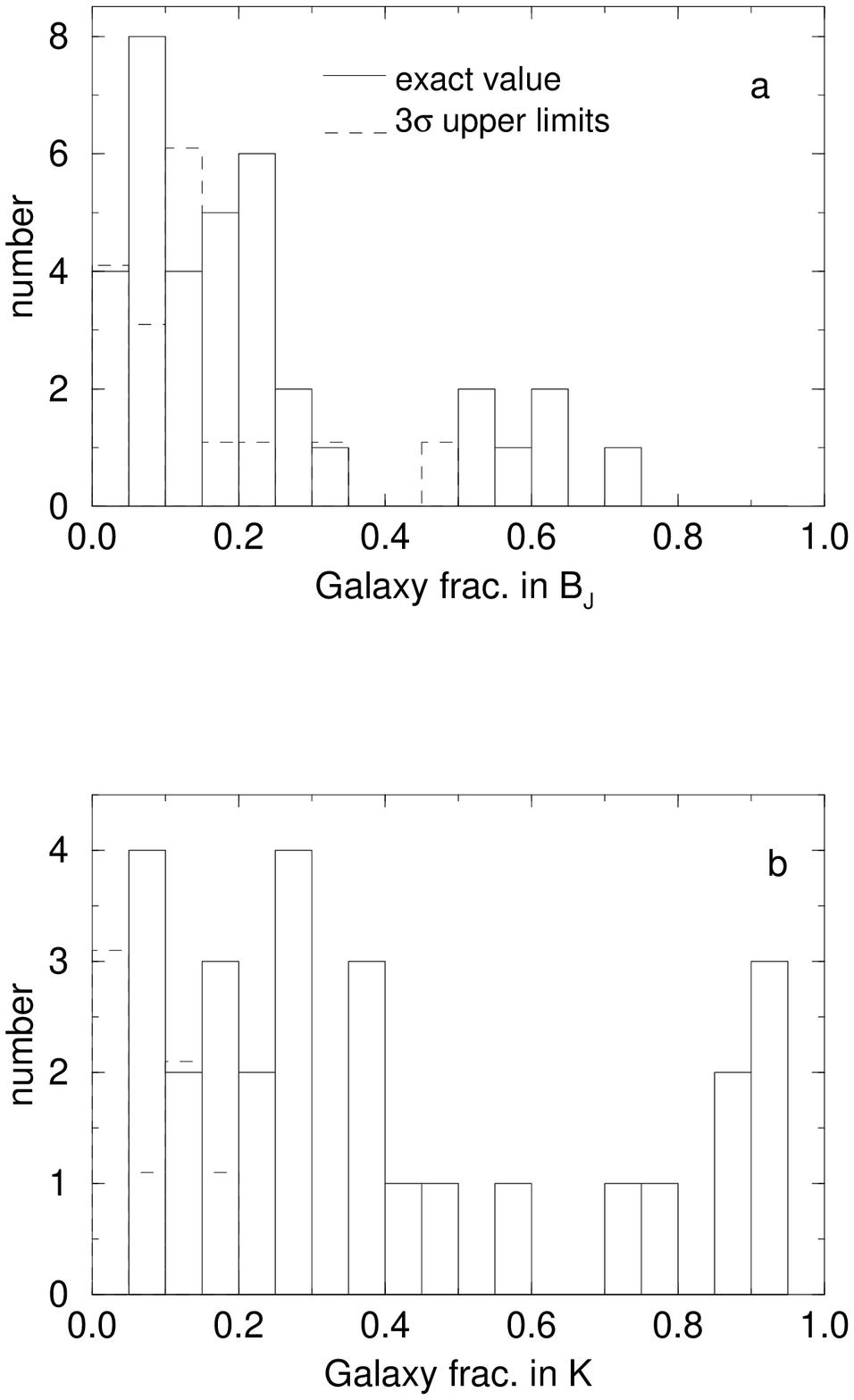}
\caption[]
{(a) Distribution in fractional galaxy contributions
in $B_{J}$
and (b) $K$
for $z<1$.
Dashed portions represent $3\sigma$ upper limits,
(see section~\ref{galcont}).}
\label{fracBK}
\end{figure}

The distributions in galaxy fractions in $B_{J}$ and $K$ are shown in
Figs.~\ref{fracBK}a and b respectively.
Sources with mean galaxy fractions 
$\simlt5\%$ are replaced by their
$3\sigma$ upper limits (dashed histograms). 
The distributions in Figs.~\ref{fracBK}a and b appear very similar, except
however for a greater number of sources
with galaxy contributions $>70\%$ in the $K$-band. 
These are all low redshift sources with strong 4000\AA$\,$ 
breaks in their
spectra. Their light is expected to be dominated
by evolved stellar populations, and hence
strongest in $K$. 
We must also note that the $B_{J}$ magnitudes
used to estimate the $K$ spectral fluxes via 
$B_{J}-K$ colours
are only accurate to
$\sim1$ mag~\cite{Drinkwater1997}. 
This photometric uncertainty is likely 
to contribute significant scatter  
in our estimates of the $K$ galaxy fraction in Fig.~\ref{fracBK}b. 
 
\begin{figure}
\vspace{-0.7in}
\plotonesmall{1.1}{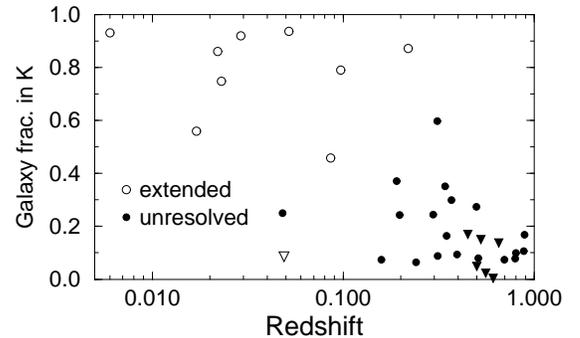}
\vspace{-1.6in}
\caption[]{Fractional
galaxy contribution in $K$ as a function of
redshift for resolved sources
(extended on $K$ and $B_{J}$ images; open symbols) and
unresolved sources (closed symbols). Triangles represent $3\sigma$ upper limits
on the galaxy fraction.}
\label{Kfracvsz}
\end{figure}

We have divided the $K$-band identifications into two populations:
those which show extended (resolved) structure, and those which remain
unresolved.
The galaxy fraction in $K$ 
for these two populations is shown as a function of $z$
in Fig.~\ref{Kfracvsz}.
Open symbols represent resolved sources, and filled symbols, unresolved
sources.
As expected, those sources exhibiting resolved structure are also those
which show large galaxy fractions and are at relatively low redshifts. 

We now investigate whether emission from the host 
galaxies of Parkes quasars can 
significantly contribute to their observed $B_{J}-K$ colours. 
We do this by computing the $B_{J}-K$ colour of the hypothesised
underlying ``quasar'', $(B_{J}-K)_{q}$,  
we would expect if contribution from the host galaxy was absent in each source.
If the observed colours were entirely due to galactic emission, then we
expect the distribution in $(B_{J}-K)_{q}$ to show a relatively small
scatter, i.e. similar to that observed for 
optically-selected quasars where typically 
$(B_{J}-K)_{q}\simeq2.5$.
 
\begin{figure}
\vspace{-0.1in}
\plotonesmall{1.1}{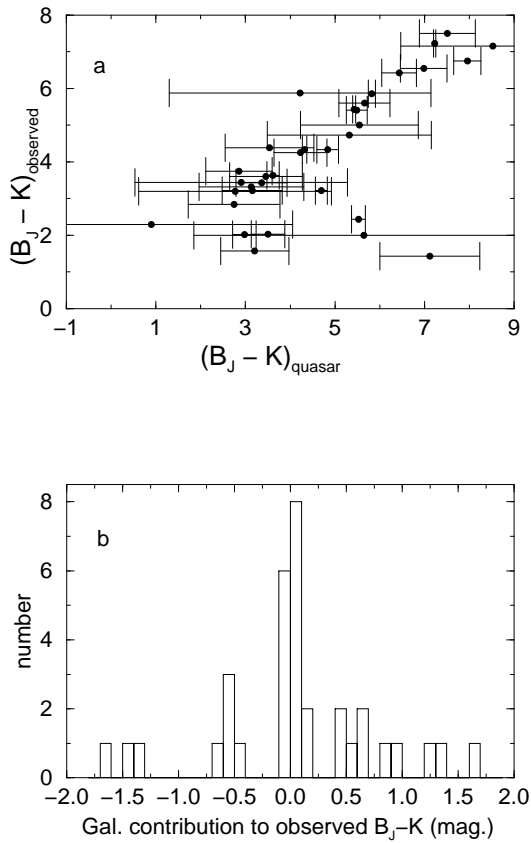}
\caption[]{(a) Observed
$B_{J}-K$ colour vs. colour of
the underlying quasar (i.e. resulting galaxy subtracted colour:
$(B_{J}-K)_{q}$). Error bars correspond to $1\sigma$ significance.
(b) Distribution in galaxy contribution to
observed $B_{J}-K$ colours in magnitudes.}
\label{BminusKgal}
\end{figure}

The colour of an underlying quasar, $(B_{J}-K)_{q}$, can be written in terms
of the observed colour $(B_{J}-K)_{obs}$ and the galaxy
fractional contributions $F_{gal}(B_{J})$ and $F_{gal}(K)$ as follows: 
\begin{equation}
(B_{J}-K)_{q}\,=\,(B_{J}-K)_{obs}\,+\,2.5\log{\left[\frac{1-F_{gal}(K)}
{1-F_{gal}(B_{J})}\right]}.  
\label{BKq}
\end{equation}
$(B_{J}-K)_{q}$ is plotted against $(B_{J}-K)_{obs}$ in Fig.~\ref{BminusKgal}a. As can be seen, the scatter in galaxy subtracted colours,
$(B_{J}-K)_{q}$, remains and is extremely similar to that of
the observed colour distribution. 
We quantify the galaxy contribution
to the observed $B_{J}-K$ colours (in magnitudes)
in Fig.~\ref{BminusKgal}b. 
From Figs.~\ref{BminusKgal}a and b, we conclude that the observed spread
in colours cannot be due to emission from the host galaxies of Parkes
quasars. An independent mechanism
must be involved.
 
\section{A Test for the Unified Model}
\label{tuf}

Studies of the host galaxy properties of BL-Lacs and radio-quasars
can be used as a test of the unified scheme for radio-loud AGN. 
Motivated by the 
canonical axisymmetric model for AGN,
the basis of this scheme is 
that the appearence of an extragalactic radio source 
is primarily determined
by viewing geometry.
Extended FRI and FRII-type radio galaxies are believed
to represent the parent (misaligned) populations of 
the more compact BL-Lacs and radio quasars
respectively~\cite{Urry1995}.
If classification is purely based on orientation, 
then intrinsic properties such as 
host galaxy luminosity should  
be approximately uniform throughout. 
In this section, we shall test this hypothesis.
 
Studies have shown that for quasars and 
FRII radio galaxies at redshifts $z\simlt0.3$, 
the situation is not
entirely clear. 
From comparisons of their mean host galaxy luminosities,
some studies have shown that FRII hosts are fainter by $\sim0.5$-1 mag. 
\cite{Smith1989}, while others have concluded that they
are comparable (eg. Taylor et al. 1996).
The main difficulty in these studies was finding sufficiently large
samples of radio galaxies and quasars matched both in 
radio power and redshift.
There is strong observational evidence however that the low redshift
BL-Lacs reside in giant ellipticals with mean optical
luminosities and de Vaucoleurs $r^{1/4}$ law profiles similar to those in
FRIs~\cite{Ulrich1988,Stichel1993}.
Very little is known about the host galaxies
of compact radio sources at higher redshifts.
Using our algorithm however, we can get estimates
of host galaxy $K$ magnitudes for sources up to $z\sim1$. 

We can predict the host galaxy $K$ magnitudes of Parkes sources 
directly from our estimates of the fractional galaxy contribution in the
$K$-band, $F_{gal}(K)$ (see section~\ref{galcont}). 
The host galaxy magnitude in an observer's frame can be written:
\begin{equation}
K_{gal}\,=\,K_{source}\,-\,2.5\log{\left[F_{gal}(K)\right]},
\label{Kgalobs}
\end{equation}
where $K_{source}$ is the observed $K$ magnitude of the source.
As discussed in section~\ref{galcont}, $F_{gal}(K)$ is determined 
from the median spectral flux at $\lambda\simeq2.2\mu$m (Eqn.~\ref{galfrac}), 
which we estimate 
using the observed spectral flux at $\lambda\simeq4400$\AA$\,$
and $B_{J}-K$ colour. 
 
\begin{figure}
\vspace{-1.1in}
\plotonesmall{1.1}{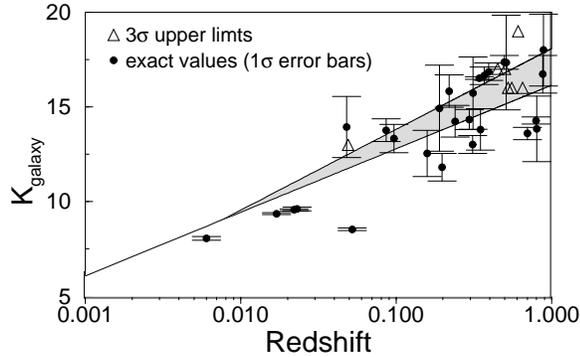}
\vspace{-1.7in}
\caption[]{Host galaxy $K$
magnitude as a function of redshift
for Parkes sources (symbols) and the range observed for radio galaxies
(shaded region; McCarthy 1993).}
\label{Kgalfig}
\end{figure}

Estimates of $K_{gal}$ as a function of redshift are shown 
in Fig.~\ref{Kgalfig}. 
For comparison, we also show the range
observed for radio galaxies (shaded region) as determined from a number of
independent studies (McCarthy 1993). 
Within our quoted uncertainties, there appears to be no significant
difference in the mean host galaxy magnitude of 
``compact'' Parkes sources and extended radio galaxies at the
redshifts indicated. 
The compact sources however appear to show a larger scatter in $K_{gal}$
at some redshift.
Since the $K_{gal}$ values were determined from
non-contemporaneous measures of observed $B_{J}$ spectral fluxes and
$B_{J}-K$ colours, this may be attributed to variability in the underlying AGN.
We are unable at present to quantify this uncertainty.
From Fig.~\ref{Kgalfig}, we conclude that the host galaxy 
luminosities of these two classes
of radio source is consistent with that required by the unified model. 

\section{Discussion}
\label{diseven}

Our results of section~\ref{galcont} clearly show that galactic emission
is unlikely to fully explain
the dispersion in $B_{J}-K$ colours observed.
This conclusion is only valid however for sources at redshifts $z\simlt 0.9$.
At higher redshifts, the 4000\AA$\,$
break feature on which our algorithm is based
is redshifted out of our observational wavelength range.
Taking into account our completeness in spectral data (section~\ref{data}), 
only 53 of the 323 
sources in the Drinkwater et al.~\shortcite{Drinkwater1997} 
sample have been analysed using our
algorithm. Since we are limited to redshifts $z\simlt 0.9$, it is 
possible that we are biased towards detecting relatively large 
galaxy contributions. Our results may thus not be representative for
the whole sample of Parkes quasars. 
 
We have two strong pieces of observational evidence that supports 
a minimal galaxy contribution from the high redshift Parkes sources: 
First, all sources with $z\simgt0.5$ appear very 
compact in $K$ (eg. Fig.~\ref{Kfracvsz}), 
and exhibit broad-line equivalent widths
typical of those
observed in optically-selected quasars.
Second, significantly high levels of linear polarisation ($\simgt5\%$)
have been observed in the near-IR in several sources at $z\simgt1$
(see Masci 1997).
This strongly indicates 
that the emission is dominated by a non-thermal mechanism. It is
important to note however that the number of sources in which
a polarisation has been searched for is 
too low to draw any reasonable conclusion. 
Further polarimetric studies of preferably the reddest
quasars 
are necessary to asses the importance of a host galaxy component. 

The next significant step in improving the algorithm presented here 
for determining the
galaxy contribution would be to completely discard 
our assumption of a power-law 
for the
shape of the underlying quasar spectrum. Although a simple power-law
is a good overall representation for the {\it continuum} around 
$\lambda\simeq4000$\AA, generic quasar spectra also include a complex
blend of emission line features superimposed on the 
continuum about this region.
As seen in most 
compilations of composite QSO spectra (eg. Francis et al. 1991),
wavelengths shortwards of $3700$\AA~ are contaminated by a Balmer coninuum, where
emission from the convergence of high order Balmer lines can introduce a steep rise in
this part of the spectrum. Furthermore there is also a complex blend of
FeII emission setting in at these wavelengths.
These considerations therefore invalidate
our assumption of a pure power-law
for the ``smooth'' quasar spectrum.
Nonetheless, it is likely that
the omission of these additional intrinsic features at $\lambda\simlt4000$\AA~
will have led us to
overestimate the galaxy contribution on average. 

\section{Conclusions}
\label{concsix}
 
This paper has explored whether emission from the host galaxies
of Parkes quasars can significantly
contribute to the relatively large spread in $B_{J}-K$ colours observed. 
If the hosts are classical giant ellipticals and their flux strongly 
contributes, then this would be expected since elliptical colours 
are known to be quite red in $B-K$ to $z\sim2$.
 
We have devised an algorithm that measures the
relative galaxy contribution in each source in an unbiased way using
the characteristic 4000\AA$\,$ break feature of elliptical galaxy SEDs.
The basis of the algorithm involves subtracting a generic elliptical SED  
from each source spectrum until the 4000\AA$\,$ break feature disapears 
and what is left is a ``smooth'' spectrum containing no breaks. 
This ``smooth'' spectrum we refer to as the underlying quasar spectrum.
The only requirement by our algorithm is that this remaining 
spectrum be smooth.
The galactic 
contribution, relative to the total light at any wavelength 
is estimated from the amount of galaxy subtracted.

The main conclusions are:
\\\indent 1. For $z\simlt0.9$, (for which the 4000\AA$\,$ feature remains
observable in our spectra),
we find broad and almost bimodal distributions in the 
relative galaxy fraction in $B_{J}$ and $K$.
Most sources
($\simgt70\%$) have galaxy
fractions $<0.3$ at the $3\sigma$ level in both
$B_{J}$ and $K$. 
The remainder have large galaxy contributions and are
predominately low redshift galaxies with strong 4000\AA$\,$ breaks. 
All of these latter sources are spatially extended and resolved 
on $B_{J}$ and $K$-band images.
In particular, there is a clear distinction in the strength of the
4000\AA$\,$ break for resolved and unresolved sources. 
\\\indent 2. Using these estimates, we find that the mean
$K$-band magnitude of the host galaxies of flat spectrum radio quasars
is consistent with that of extended radio galaxies at $z\simlt0.9$. 
This is consistent with the unified model for radio-loud
AGN.
\\\indent 3. By subtracting the galaxy contribution in each 
bandpass from the observed $B_{J}-K$ colours 
of Parkes sources, we find that at predominately 
the $2\sigma$ confidence level, 
the relatively large spread in colours still 
remains. 
We conclude that in a majority of cases, the 
relatively red colours
must be due to a mechanism
other than that contributed by a ``red'' stellar component.

\section{Acknowledgments}
FJM acknowledges support from an APRA Scholarship, and RLW
from an ARC research grant.

\end{document}